# BO Ari Light Curve Analysis using Ground-Based and TESS Data

Atila Poro[1], Shiva Zamanpour[1], Maryam Hashemi[1], Yasemin Aladağ[3], Nazim Aksaker[2,3], Samaneh Rezaei[1], Arif Solmaz[3]

[1]The International Occultation Timing Association Middle East section, Iran, info@iota-me.com
[2]Adana Organised Industrial Zones Vocational School of Technical Science, University of Çukurova, 01410, Adana, Turkey
[3]Space Science and Solar Energy Research and Application Center (UZAYMER), University of Çukurova, 01330, Adana, Turkey

**ABSTRACT**

We present new BVR band photometric light curves of BO Aries obtained in 2020 and combined them with the Transiting Exoplanet Survey Satellite (TESS) light curves. We obtained times of minima based on Gaussian and Cauchy distributions and then applied the Monte Carlo Markov Chain (MCMC) method to measure the amount of uncertainty from our CCD photometry and TESS data. A new ephemeris of the binary system was computed employing 204 times of minimum. The light curves were analyzed using the Wilson-Devinney binary code combined with the Monte Carlo (MC) simulation. For this light curve solution, we considered a dark spot on the primary component. We conclude that this binary is an A-type system with a mass ratio of $q = 0.2074 \pm 0.0001$, an orbital inclination of $i = 82.18 \pm 0.02$ deg, and a fillout factor of $f = 75.7 \pm 0.8\%$. Our results for the $a$ ($R_\odot$) and $q$ parameters are consistent with the results of the Xu-Dong Zhang and Sheng-Bang Qian (2020) model. The absolute parameters of the two components were calculated and the distance estimate of the binary system was found to be $142 \pm 9$ pc.

## 1. INTRODUCTION

EW-type binaries are systems with main characteristics such as fairly equal depth eclipsing light curves, short orbital periods, less than one day, and frequently mass transfer between the two components which are in contact with each other sharing a common convective envelope. Eclipsing binaries can provide fundamental stellar properties and critical tests on the theories of stellar evolution and structure (Yuan et al. 2019).

Nicholson and Varley (2006) discovered the short-period binary system BO Ari (ASAS J021208+2708.2) and determined the first orbital period of $0.3182^d$. It was later observed by Acerbi et al. (2011) and classified as an A-type W Ursae Majoris system with a mass ratio of $q = 0.1889$ and contact degree of $f = 58.7\%$. Gürol et al. (2015) made the first Spectroscopic observations and derived the $q = 0.19024$ and $f = 49.8\%$ by combining a light and radial velocity curve solution. Through the $BVR$ observations done by Kriwattanawong et al. (2016) a mass ratio of $q = 0.1754$ and contact degree of $f = 27.72\%$ were obtained for the binary system.

In this study, the multi-color CCD light curves in $B$, $V$, and $R$ bands along with photometric data obtained from the TESS are presented. We determined a new ephemeris for this binary system. The light curve solution with Wilson-Devinney code combined with the MC simulation was performed to obtain reliable photometric parameters. Absolute parameters and distance of the system were derived.

## 2. NEW PHOTOMETRIC OBSERVATION

The observation of BO Ari was carried out by a 50 cm Ritchey-Chretien RC 500/4000 Pro RC SGA OTA telescope and Apogee Aspen CG6 LN-2-G07-S58 type CCD during nine nights of observation at the UZAYMER Observatory, Çukurova University, Adana, Turkey in January 2020. The CCD has a $1024 \times 1024$ pixel array with a pixel length of 24µ. In these observation nights, we used the $BVR$ standard Johnson filters for the photometry. Each of the frames was $2 \times 2$ binned with 40s exposure time for $R$ filter, 60s for $V$ filter, and 90s for $B$ filter; the average temperature of the CCD was -43°C during the observations.

TYC 1761-2002-1, TYC 1761-1358-1, WISEA J021218.9 (NED), and WISEA J021158.5 (NED) were chosen as comparison stars, and BD+26 369 was selected as a check star. All of these stars are close to BO Ari and the magnitude of the check star is appropriate. Figure 1 shows an observed field-of-view for BO Ari with the comparisons and check stars; The characteristics of these stars are shown in Table 1.

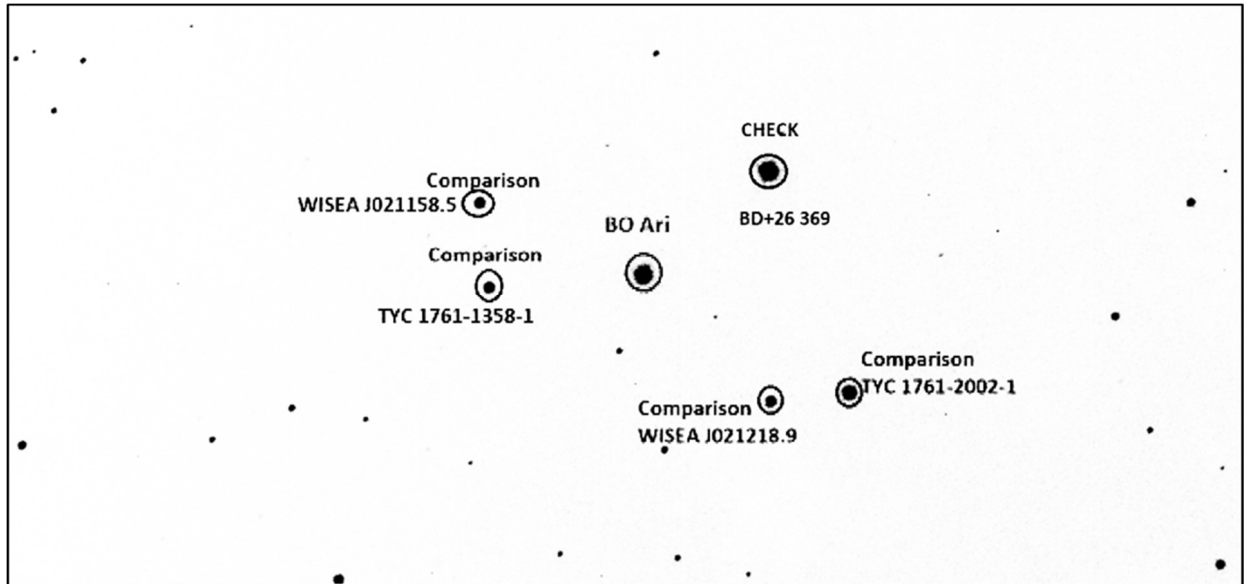

Figure 1. Observed field-of-view for BO Ari, comparison stars, and check star.

Table 1. Characteristics of the variable, comparison and check stars (from: Vizier-APASS9).

| Star type | Star name | RA (J2000) | DEC (J2000) | Magnitude ($V$) |
|---|---|---|---|---|
| Variable | BO Ari | 02 12 08.7070 | +27 08 18.6180 | 10.169 |
| Check | BD+26 369 | 02 12 11.1814 | +27 10 19.8948 | 10.263 |
| Comparison | TYC 1761-2002-1 | 02 12 22.4090 | +27 08 33.4968 | 11.588 |
| Comparison | TYC 1761-1358-1 | 02 12 01.7561 | +27 07 03.1260 | 12.061 |
| Comparison | WISEA J021218.9 (NED) | 02 12 19.0166 | +27 07 52.9630 | 12.780 |
| Comparison | WISEA J021158.5 (NED) | 02 11 58.5466 | +27 07 53.1830 | 12.590 |

We reduced the raw CCD images and the basic data reduction was performed for bias, dark and flat-field according to the standard method. We aligned, reduced, and plotted raw images with AstroImageJ (AIJ) software (Collins et al. 2017). AIJ provides an astronomy-specific image display environment and tools for astronomy-specific image calibration and data reduction. This software has been determined to identify the best linear fit of a dataset (by exerting air mass) to the light curve (Davoudi et al. 2020).

The study of eclipsing binaries has been advanced significantly with an increasing rate of discoveries as a result of space satellites. The Transiting Exoplanet Survey Satellite (TESS) is one of the recent missions that obtained new photometric observations data from numerous numbers of known eclipsing binaries. In this mission, the studied stars were 30 to 100 times brighter than those the Kepler mission and K2 follow-up surveyed, which enabled far easier follow-up observations with both ground-based and space-based telescopes. TESS also covered a sky area 400 times larger than that monitored by Kepler.

BO Ari (TIC 5674169) was observed by the TESS mission. TESS data of this binary system is available at the Mikulski Space Telescope Archive (MAST). We extracted TESS style curves using LightKurve code[1] from the MAST. It was observed in sector 18 by Camera 1 and CCD 3 in 120 seconds cadences. Its detrended light curves were extracted from the MAST.

## 3. NEW EPHEMERIS

---
[1]https://docs.lightkurve.org/

Four primary and five secondary minimum times were determined from our observed light curves. They were found through fitting the models to the light curves and existing minima based on Gaussian and Cauchy distributions. We then employed the Monte Carlo Markov Chain (MCMC) method to measure the amount of uncertainty related to each value (Poro et al. 2020). We also benefited from Python along with its PyMC3 package to execute the lines of code (Salvatier et al. 2016). According to this method, we also calculated all times of minima of the TESS data for this binary system. Thus, 48 new times of minima were obtained from the TESS data.

Based on our observations, TESS data, and other literature we obtained 204 timings of minimum light including 98 primary and 106 secondary times of minima, given in Table 2.

We used the following reference ephemeris (Acerbi et al. 2011), for calculating epoch and O-C values,

$$BJD_{TDB}\,(Min.I) = 2452625.64163 + 0.3181963 \times E. \quad (1)$$

We fitted all mid-transit timings with a line, using the Robust regression and determined a new ephemeris for primary minimum as,

$$Min\,I\,(BJD_{TDB}) = (2452625.650338 \pm 0.001993) + (0.318194081 \pm 0.000000114) \times E\,days \quad (2)$$

where $E$ is an integer of orbital cycles after the reference epoch.

**Table 2. Times of minima of BO Ari.**

| BJD$_{TDB}$ (Min.) | Error | Epoch | O-C | Reference | BJD$_{TDB}$ (Min.) | Error | Epoch | O-C | Reference |
|---|---|---|---|---|---|---|---|---|---|
| 2451479.6573 | | -3601.5 | -0.0004 | Nicholson et al. 2006 | 2458798.7735 | 0.0002 | 19400.5 | -0.0354 | TESS |
| 2452625.6416 | | 0 | 0 | Acerbi et al. 2011 | 2458798.9341 | 0.0001 | 19401 | -0.0339 | TESS |
| 2454049.4086 | 0.0025 | 4474.5 | -0.0024 | Martignoni 2011 | 2458799.0918 | 0.0002 | 19401.5 | -0.0353 | TESS |
| 2454062.4588 | | 4515.5 | 0.0018 | Acerbi et al. 2011 | 2458799.2523 | 0.0001 | 19402 | -0.0339 | TESS |
| 2454080.4341 | | 4572 | -0.0010 | Acerbi et al. 2011 | 2458799.4099 | 0.0002 | 19402.5 | -0.0354 | TESS |
| 2454081.3885 | | 4575 | -0.0012 | Acerbi et al. 2011 | 2458799.5706 | 0.0001 | 19403 | -0.0338 | TESS |
| 2454083.2986 | | 4581 | -0.0003 | Acerbi et al. 2011 | 2458799.7281 | 0.0002 | 19403.5 | -0.0354 | TESS |
| 2454083.4591 | | 4581.5 | 0.0011 | Acerbi et al. 2011 | 2458799.8888 | 0.0001 | 19404 | -0.0338 | TESS |
| 2454084.4135 | 0.0003 | 4584.5 | 0.0009 | Martignoni 2011 | 2458800.0463 | 0.0002 | 19404.5 | -0.0354 | TESS |
| 2454095.3901 | | 4619 | -0.0002 | Acerbi et al. 2011 | 2458800.2071 | 0.0001 | 19405 | -0.0337 | TESS |
| 2454507.2988 | 0.0100 | 5913.5 | 0.0033 | Paschke 2009 | 2458800.3645 | 0.0002 | 19405.5 | -0.0354 | TESS |
| 2454808.3048 | 0.0100 | 6859.5 | -0.0043 | Paschke 2009 | 2458800.5252 | 0.0001 | 19406 | -0.0338 | TESS |
| 2455080.3633 | 0.0003 | 7714.5 | -0.0037 | Demircan et al. 2011 | 2458800.6827 | 0.0002 | 19406.5 | -0.0354 | TESS |
| 2455080.5190 | 0.0001 | 7715 | -0.0071 | Gokay et al. 2010 | 2458800.8435 | 0.0001 | 19407 | -0.0337 | TESS |
| 2455103.4287 | 0.0003 | 7787 | -0.0075 | Gokay et al. 2010 | 2458801.1164 | 0.0021 | 19408 | -0.0790 | TESS |
| 2455103.5895 | 0.0004 | 7787.5 | -0.0058 | Gokay et al. 2010 | 2458801.0023 | 0.0003 | 19407.5 | -0.0340 | TESS |
| 2455144.3197 | 0.0008 | 7915.5 | -0.0047 | Gokay et al. 2010 | 2458801.6384 | 0.0002 | 19409.5 | -0.0343 | TESS |
| 2455392.5079 | 0.0001 | 8695.5 | -0.0097 | Demircan et al. 2011 | 2458801.7980 | 0.0001 | 19410 | -0.0338 | TESS |
| 2455487.3302 | 0.0002 | 8993.5 | -0.0099 | Gokay et al. 2012 | 2458801.9552 | 0.0002 | 19410.5 | -0.0357 | TESS |
| 2455487.4897 | 0.0001 | 8994 | -0.0095 | Gokay et al. 2012 | 2458802.1158 | 0.0002 | 19411 | -0.0342 | TESS |
| 2455509.2849 | | 9062.5 | -0.0107 | Gürol et al. 2015 | 2458803.5463 | 0.0002 | 19415.5 | -0.0356 | TESS |
| 2455509.4448 | 0.0008 | 9063 | -0.0099 | Gokay et al. 2012 | 2458803.7071 | 0.0001 | 19416 | -0.0339 | TESS |
| 2455528.3776 | | 9122.5 | -0.0098 | Gokay et al. 2012 | 2458803.8645 | 0.0002 | 19416.5 | -0.0356 | TESS |
| 2455538.2425 | | 9153.5 | -0.0090 | Gürol et al. 2015 | 2458804.0254 | 0.0001 | 19417 | -0.0338 | TESS |
| 2455551.2880 | 0.0001 | 9194.5 | -0.0095 | Gokay et al. 2012 | 2458804.1827 | 0.0002 | 19417.5 | -0.0356 | TESS |
| 2455553.1970 | 0.0001 | 9200.5 | -0.0097 | Gokay et al. 2012 | 2458804.3435 | 0.0001 | 19418 | -0.0339 | TESS |
| 2455553.3538 | | 9201 | -0.0120 | Gürol et al. 2015 | 2458804.5009 | 0.0002 | 19418.5 | -0.0356 | TESS |
| 2455887.1404 | | 10250 | -0.0133 | Kriwattanawong 2016 | 2458804.6616 | 0.0001 | 19419 | -0.0340 | TESS |
| 2455890.1630 | | 10259.5 | -0.0136 | Kriwattanawong 2016 | 2458804.8192 | 0.0002 | 19419.5 | -0.0355 | TESS |
| 2455890.3228 | | 10260 | -0.0129 | Kriwattanawong 2016 | 2458804.9797 | 0.0001 | 19420 | -0.0341 | TESS |
| 2455892.0724 | | 10265.5 | -0.0133 | Kriwattanawong 2016 | 2458805.1373 | 0.0002 | 19420.5 | -0.0356 | TESS |
| 2455892.2325 | | 10266 | -0.0123 | Kriwattanawong 2016 | 2458805.2979 | 0.0001 | 19421 | -0.0341 | TESS |
| 2456167.9468 | 0.0020 | 11132.5 | -0.0151 | Nelson 2013 | 2458805.4556 | 0.0002 | 19421.5 | -0.0355 | TESS |
| 2456279.6330 | 0.0001 | 11483.5 | -0.0158 | Diethelm 2013 | 2458805.6159 | 0.0001 | 19422 | -0.0343 | TESS |
| 2456514.9366 | 0.0002 | 12223 | -0.0184 | Nelson 2014 | 2458805.7738 | 0.0002 | 19422.5 | -0.0355 | TESS |
| 2456516.5268 | | 12228 | -0.0192 | Paschke 2014 | 2458805.9342 | 0.0001 | 19423 | -0.0342 | TESS |
| 2456526.8689 | 0.0003 | 12260.5 | -0.0185 | Nelson 2014 | 2458806.0920 | 0.0002 | 19423.5 | -0.0355 | TESS |
| 2456535.9372 | 0.0002 | 12289 | -0.0188 | Nelson 2014 | 2458806.2523 | 0.0001 | 19424 | -0.0343 | TESS |
| 2456560.9153 | | 12367.5 | -0.0191 | Nelson 2014 | 2458806.4102 | 0.0002 | 19424.5 | -0.0355 | TESS |
| 2456579.6893 | 0.0002 | 12426.5 | -0.0187 | Nelson 2014 | 2458806.5705 | 0.0001 | 19425 | -0.0343 | TESS |
| 2456579.8483 | 0.0002 | 12427 | -0.0188 | Nelson 2014 | 2458806.7284 | 0.0002 | 19425.5 | -0.0355 | TESS |
| 2456630.2818 | 0.0025 | 12585.5 | -0.0194 | Hubscher 2014 | 2458806.8887 | 0.0001 | 19426 | -0.0343 | TESS |
| 2456630.4414 | 0.0018 | 12586 | -0.0189 | Hubscher 2014 | 2458807.0465 | 0.0002 | 19426.5 | -0.0356 | TESS |
| 2456928.7469 | 0.0002 | 13523.5 | -0.0224 | Nelson 2015 | 2458807.2069 | 0.0001 | 19427 | -0.0343 | TESS |
| 2456949.1106 | | 13587.5 | -0.0233 | Nagai 2015 | 2458807.3647 | 0.0002 | 19427.5 | -0.0355 | TESS |
| 2456949.2708 | | 13588 | -0.0222 | Nagai 2015 | 2458807.5250 | 0.0001 | 19428 | -0.0343 | TESS |
| 2456978.3858 | 0.0014 | 13679.5 | -0.0221 | Hubscher 2015 | 2458807.6829 | 0.0002 | 19428.5 | -0.0355 | TESS |
| 2456978.5435 | 0.0015 | 13680 | -0.0235 | Hubscher 2015 | 2458807.8432 | 0.0001 | 19429 | -0.0343 | TESS |
| 2457329.3525 | | 14782.5 | -0.0259 | Hubscher 2017 | 2458808.0012 | 0.0002 | 19429.5 | -0.0354 | TESS |
| 2457329.5109 | | 14783 | -0.0266 | Hubscher 2017 | 2458808.1614 | 0.0001 | 19430 | -0.0343 | TESS |
| 2457329.6699 | | 14783.5 | -0.0267 | Hubscher 2017 | 2458808.3193 | 0.0002 | 19430.5 | -0.0355 | TESS |
| 2457758.2768 | | 16130.5 | -0.0302 | Paschke 2017 | 2458808.4796 | 0.0001 | 19431 | -0.0343 | TESS |
| 2458098.5868 | | 17200 | -0.0312 | Nelson 2018 | 2458808.6375 | 0.0002 | 19431.5 | -0.0355 | TESS |
| 2458380.5070 | 0.0001 | 18086 | -0.0329 | Ozavci et al. 2019 | 2458808.7977 | 0.0001 | 19432 | -0.0344 | TESS |
| 2458790.8188 | 0.00016 | 19375.5 | -0.0352 | TESS | 2458808.9558 | 0.0002 | 19432.5 | -0.0354 | TESS |
| 2458790.9789 | 0.00013 | 19376 | -0.0342 | TESS | 2458809.1159 | 0.0001 | 19433 | -0.0344 | TESS |
| 2458791.4550 | 0.00017 | 19377.5 | -0.0354 | TESS | 2458809.2739 | 0.0002 | 19433.5 | -0.0355 | TESS |
| 2458791.6154 | 0.00013 | 19378 | -0.0341 | TESS | 2458809.4340 | 0.0001 | 19434 | -0.0345 | TESS |
| 2458791.7733 | 0.00017 | 19378.5 | -0.0353 | TESS | 2458809.5923 | 0.0002 | 19434.5 | -0.0353 | TESS |
| 2458791.9336 | 0.00013 | 19379 | -0.0341 | TESS | 2458809.7521 | 0.0001 | 19435 | -0.0346 | TESS |
| 2458792.0915 | 0.00017 | 19379.5 | -0.0353 | TESS | 2458809.9105 | 0.0002 | 19435.5 | -0.0353 | TESS |
| 2458792.2518 | 0.00012 | 19380 | -0.0341 | TESS | 2458810.0702 | 0.0001 | 19436 | -0.0347 | TESS |
| 2458792.4097 | 0.00017 | 19380.5 | -0.0353 | TESS | 2458810.2288 | 0.0002 | 19436.5 | -0.0352 | TESS |
| 2458792.5700 | 0.00013 | 19381 | -0.0341 | TESS | 2458810.3883 | 0.0001 | 19437 | -0.0348 | TESS |
| 2458792.7278 | 0.00017 | 19381.5 | -0.0354 | TESS | 2458810.5470 | 0.0002 | 19437.5 | -0.0352 | TESS |
| 2458792.8883 | 0.00013 | 19382 | -0.0340 | TESS | 2458810.7065 | 0.0001 | 19438 | -0.0348 | TESS |
| 2458793.0461 | 0.00017 | 19382.5 | -0.0353 | TESS | 2458810.8653 | 0.0002 | 19438.5 | -0.0351 | TESS |
| 2458793.2064 | 0.00013 | 19383 | -0.0341 | TESS | 2458811.0246 | 0.0001 | 19439 | -0.0349 | TESS |
| 2458793.3643 | 0.00017 | 19383.5 | -0.0353 | TESS | 2458811.1836 | 0.0002 | 19439.5 | -0.0350 | TESS |
| 2458793.5246 | 0.00013 | 19384 | -0.0341 | TESS | 2458811.3427 | 0.0001 | 19440 | -0.0350 | TESS |
| 2458793.6825 | 0.00017 | 19384.5 | -0.0353 | TESS | 2458811.5019 | 0.0002 | 19440.5 | -0.0349 | TESS |
| 2458793.8428 | 0.00013 | 19385 | -0.0341 | TESS | 2458811.6608 | 0.0001 | 19441 | -0.0351 | TESS |
| 2458794.0007 | 0.00017 | 19385.5 | -0.0353 | TESS | 2458811.8202 | 0.0002 | 19441.5 | -0.0348 | TESS |
| 2458794.1610 | 0.00013 | 19386 | -0.0341 | TESS | 2458811.9789 | 0.0001 | 19442 | -0.0352 | TESS |
| 2458794.3188 | 0.00017 | 19386.5 | -0.0354 | TESS | 2458812.1384 | 0.0002 | 19442.5 | -0.0348 | TESS |
| 2458794.4792 | 0.00013 | 19387 | -0.0341 | TESS | 2458812.2970 | 0.0001 | 19443 | -0.0353 | TESS |
| 2458794.6371 | 0.00017 | 19387.5 | -0.0353 | TESS | 2458812.4569 | 0.0002 | 19443.5 | -0.0345 | TESS |
| 2458794.7974 | 0.00013 | 19388 | -0.0341 | TESS | 2458812.6151 | 0.0001 | 19444 | -0.0354 | TESS |
| 2458794.9552 | 0.00016 | 19388.5 | -0.0354 | TESS | 2458812.7748 | 0.0002 | 19444.5 | -0.0348 | TESS |
| 2458795.1156 | 0.00013 | 19389 | -0.0341 | TESS | 2458812.9334 | 0.0001 | 19445 | -0.0353 | TESS |
| 2458795.2734 | 0.00017 | 19389.5 | -0.0354 | TESS | 2458813.0931 | 0.0002 | 19445.5 | -0.0347 | TESS |
| 2458795.4338 | 0.00013 | 19390 | -0.0341 | TESS | 2458813.2514 | 0.0001 | 19446 | -0.0355 | TESS |
| 2458795.5917 | 0.00017 | 19390.5 | -0.0353 | TESS | 2458813.4112 | 0.0002 | 19446.5 | -0.0348 | TESS |
| 2458795.7520 | 0.00013 | 19391 | -0.0341 | TESS | 2458813.5697 | 0.0001 | 19447 | -0.0354 | TESS |
| 2458795.9098 | 0.00017 | 19391.5 | -0.0354 | TESS | 2458813.7294 | 0.0002 | 19447.5 | -0.0348 | TESS |
| 2458796.0702 | 0.00013 | 19392 | -0.0341 | TESS | 2458813.8879 | 0.0001 | 19448 | -0.0354 | TESS |
| 2458796.2280 | 0.00017 | 19392.5 | -0.0354 | TESS | 2458814.0477 | 0.0002 | 19448.5 | -0.0347 | TESS |
| 2458796.3884 | 0.00013 | 19393 | -0.0341 | TESS | 2458814.2058 | 0.0001 | 19449 | -0.0357 | TESS |
| 2458796.5462 | 0.00017 | 19393.5 | -0.0354 | TESS | 2458814.3659 | 0.0002 | 19449.5 | -0.0347 | TESS |
| 2458796.7066 | 0.00013 | 19394 | -0.0341 | TESS | 2458814.5240 | 0.0001 | 19450 | -0.0357 | TESS |
| 2458796.8644 | 0.00017 | 19394.5 | -0.0354 | TESS | 2458814.6842 | 0.0002 | 19450.5 | -0.0346 | TESS |
| 2458797.0250 | 0.00013 | 19395 | -0.0339 | TESS | 2458814.8422 | 0.0001 | 19451 | -0.0357 | TESS |
| 2458797.1826 | 0.00017 | 19395.5 | -0.0354 | TESS | 2458814.9411 | 0.0012 | 19451 | 0.0632 | TESS |
| 2458797.3431 | 0.0001 | 19396 | -0.0340 | TESS | 2458863.3684 | 0.0002 | 19603.5 | -0.0344 | This study |
| 2458797.5008 | 0.0002 | 19396.5 | -0.0354 | TESS | 2458868.3005 | 0.0003 | 19619 | -0.0343 | This study |
| 2458797.6613 | 0.0001 | 19397 | -0.0340 | TESS | 2458871.1632 | 0.0005 | 19628 | -0.0354 | This study |
| 2458797.8190 | 0.0002 | 19397.5 | -0.0354 | TESS | 2458871.3220 | 0.0002 | 19628.5 | -0.0357 | This study |
| 2458797.9794 | 0.0001 | 19398 | -0.0341 | TESS | 2458873.2320 | 0.0003 | 19634.5 | -0.0349 | This study |
| 2458798.1372 | 0.0002 | 19398.5 | -0.0354 | TESS | 2458874.1871 | 0.0003 | 19637.5 | -0.0344 | This study |
| 2458798.2976 | 0.0001 | 19399 | -0.0341 | TESS | 2458874.3463 | 0.0002 | 19638 | -0.0343 | This study |
| 2458798.4554 | 0.0002 | 19399.5 | -0.0354 | TESS | 2458875.2993 | 0.0002 | 19641 | -0.0359 | This study |
| 2458798.6159 | 0.0001 | 19400 | -0.0339 | TESS | 2458880.2317 | 0.0002 | 19656.5 | -0.0355 | This study |

## 4. PHOTOMETRIC SOLUTIONS

The BVR light curves from our ground-based observations and TESS light curve were analyzed with the Wilson & Devinney (1971) code combined with Monte Carlo (MC) simulations. This method can accurately obtain parameters and their uncertainties. So we used this method to search mass ratio. The free parameters in MC simulation and their ranges are given in Table 3.

Table 3. Free parameters and searching ranges in MC Simulations.

| Parameter | Value |
|---|---|
| $i$ (deg) | 60-90 |
| $T_2$ (K) | 5000-6500 |
| $\Omega_{1,2}$ | 1.5-9 |
| $q = (m_2/m_1)$ | 0.1-6 |
| $l_1$ | 1-12 |
| Phase shift | -0.03-0.03 |
| co-latitude (deg) | 0-180 |
| Longitude (deg) | 0-360 |
| Spot radius (deg) | 1-90 |
| $T_{spot}/T_1$ | 0.7-1 |

Based on our data and after calibrating (Høg et al. 2000), we calculated $(B-V)_{BO\ Ari} = 0^m.573$. As a result, according to Eker et al. (2020), the effective temperature of the primary component was found to be 5873 K. Thus, this temperature value is close to the value as the Gaia DR2 catalog (5874 K). As shown in Figure 2, the obtained temperature from derived $(B-V)$ color for the primary component of BO Ari is also in an acceptable range with the method of Sekiguchi and Fukugita (2000).

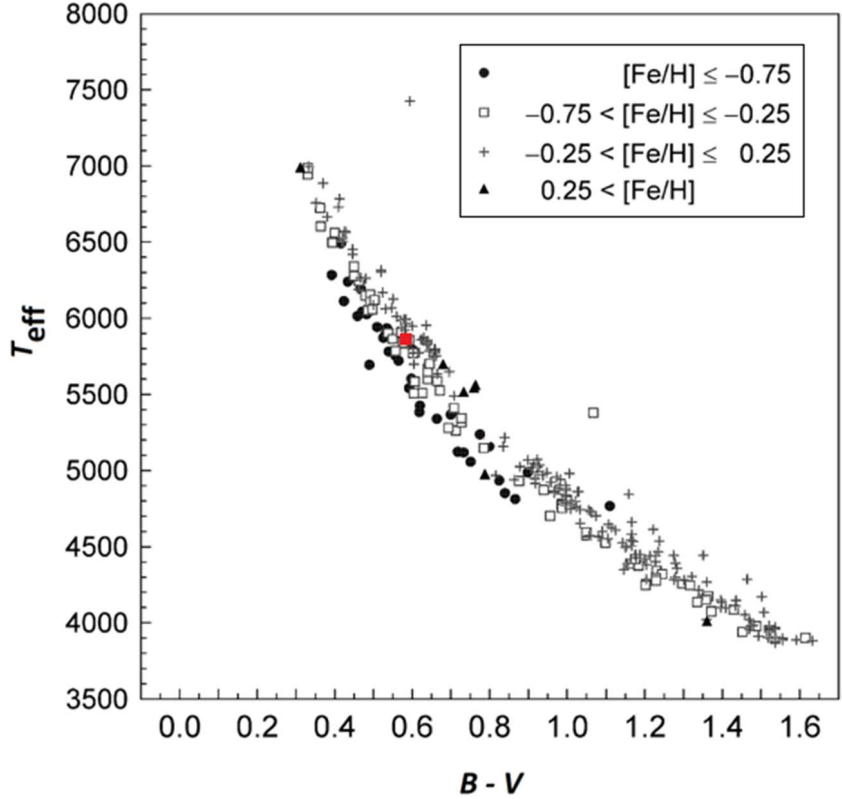

**Figure 2.** BO Ari's position (red dot) based on the Sekiguchi and Fukugita (2000) results.

The gravity-darkening coefficients $g_1 = g_2 = 0.32$ (Lucy 1967) and bolometric albedo values $A_1 = A_2 = 0.5$ (Rucinski 1969) were used, which correspond to the convective envelopes of both components. We assumed linear limb darkening coefficients taken from the tables published by Van Hamme (1993). The resulted parameter values obtained in the analysis of the light curves of BO Ari are given in Table 4, and the synthetic light curves based on these parameters are given in Figures 3 and 4.

**Table 4. Photometric solutions of BQ Ari.**

| Parameter | Results |
|---|---|
| $T_1$ (K) | 5873 |
| $T_2$ (K) | 5850(35) |
| $\Omega_1 = \Omega_2$ | 2.151(2) |
| $i$ (deg) | 82.18(2) |
| $q$ | 0.2074(1) |
| $l_1/l_{tot}(B)$ | 0.7815(6) |
| $l_2/l_{tot}(B)$ | 0.2185(4) |
| $l_1/l_{tot}(V)$ | 0.7818(6) |
| $l_2/l_{tot}(V)$ | 0.2182(4) |
| $l_1/l_{tot}(R)$ | 0.7818(6) |
| $l_2/l_{tot}(R)$ | 0.2182(4) |
| $A_1 = A_2$ | 0.50 |
| $g_1 = g_2$ | 0.32 |
| $f$ (%) | 75.7(8) |
| $r_1$(back) | 0.593(5) |
| $r_1$(side) | 0.562(4) |
| $r_1$(pole) | 0.508(4) |
| $r_2$(back) | 0.353(4) |
| $r_2$(side) | 0.278(3) |
| $r_2$(pole) | 0.263(3) |
| $r_1$(mean) | 0.553(2) |
| $r_2$(mean) | 0.295(4) |
| Colatitude$_{spot}$ (deg) | 107(2) |
| Longitude$_{spot}$ (deg) | 81(1) |
| Radius$_{spot}$ (deg) | 21(1) |
| $T_{spot}/T_{star}$ | 0.92(2) |
| Phase Shift | 0.01(1) |

Notes: Parameters of a star spot on the primary component.

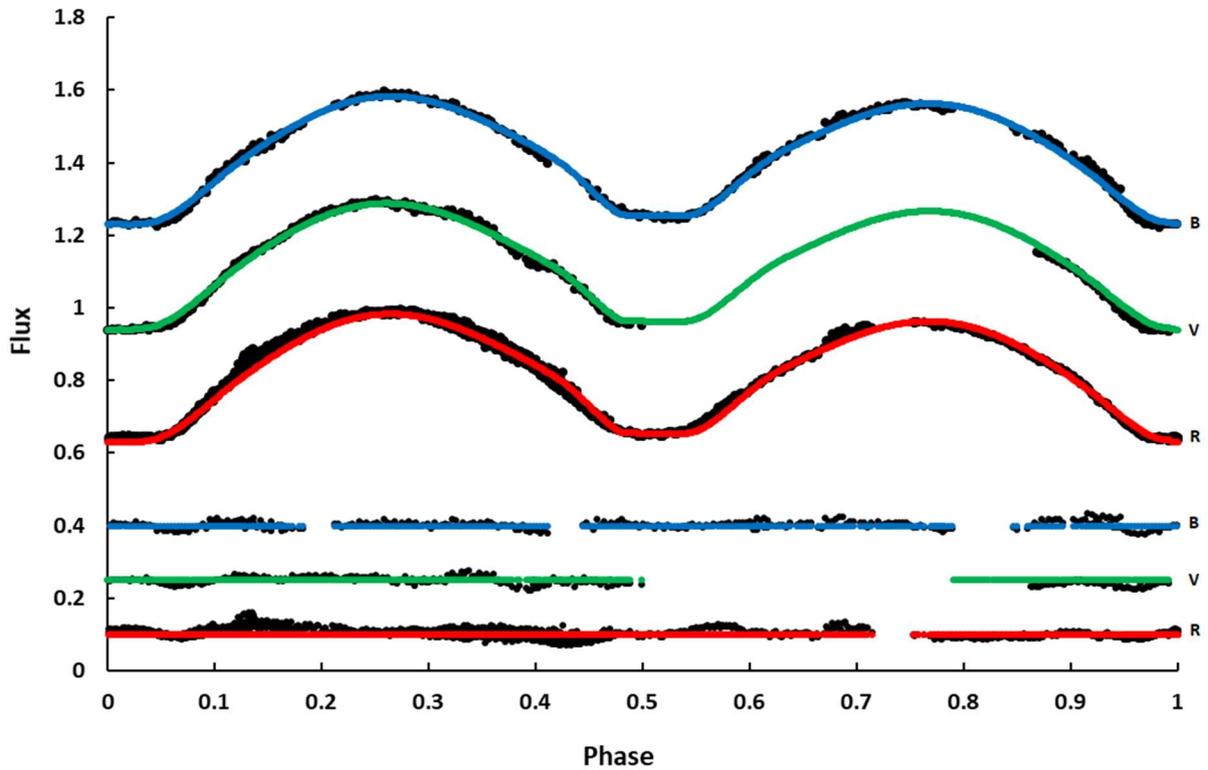

**Figure 3. The observed light curves of BO Ari (black dots), and synthetic light curves obtained from light curve solutions in the *B*, *V*, and *R* filters (top to bottom respectively) and residuals are plotted; with respect to orbital phase, shifted arbitrarily in the relative flux.**

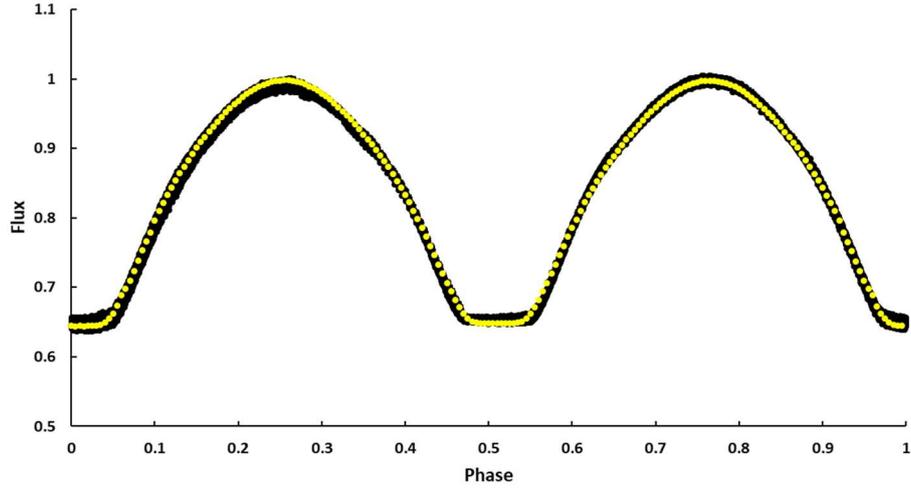

**Figure 4. TESS observation and synthetic light curves of BO Ari.**

The radial velocity of BO Ari is not available in this study, and we just can estimate the absolute parameters (Table 5). Accordingly, the mass of the primary component is derived from the Eker et al. (2020) study, and the mass of the secondary component was calculated from the equation $q = \frac{m_2}{m_1}$.

**Table 5. Estimated absolute elements of BO Ari.**

| Parameter | Primary | Secondary |
|---|---|---|
| $Mass\ (M_\odot)$ | 1.095 | 0.227(15) |
| $Radius\ (R_\odot)$ | 1.190(7) | 0.636(9) |
| $Luminosity\ (L_\odot)$ | 1.517(15) | 0.425(11) |
| $M_{bol}$ (mag) | 4.29(28) | 5.67(23) |
| $log\ g$ (cgs) | 4.326(16) | 4.187(13) |
| $a\ (R_\odot)$ | 2.152(18) | |

The 3D view and the geometrical structure of BO Ari is shown in Figure 5.

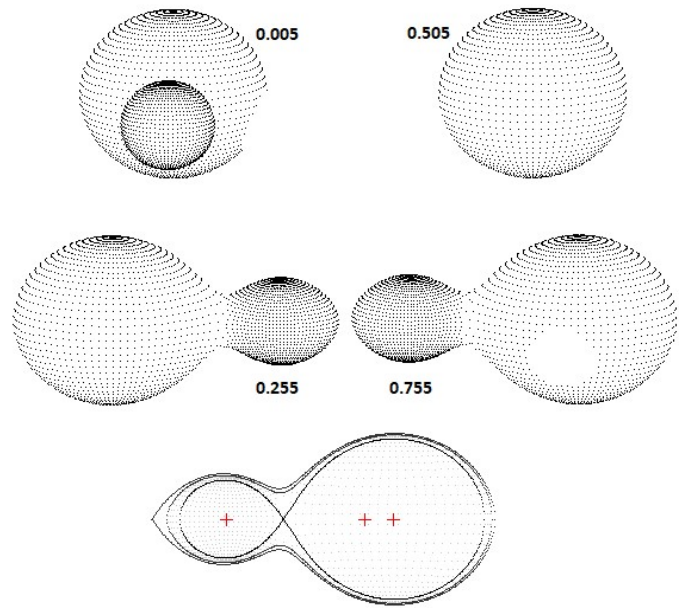

**Figure 5. The positions of the components of BO Ari.**

We estimated the binary system distance using the results of absolute parameters. The value of $m_v = 10.14 \pm 0.005$ was calculated from the observational light curve, and $M_v = 4.287 \pm 0.03$ was obtained using $BC_1 = 0.003$ according to the Eker et al. (2020). Based on these values, we calculated the binary system's distance to be $142 \pm 9$ pc, using $A_{d\,v} = 0.09 \pm 0.02$ (Schlafly and Finkbeiner 2011).

## 5. CONCLUSIONS

We obtained the temperature of the primary component as $T = 5873$ K by using $B - V$. This value is very close to the Gaia DR2's temperate for BO Ari which is $T = 5874$. It is found that BO Ari is a contact binary with a mass ratio of $q = 0.2074 \pm 0.0001$, a fillout factor of $f = 75.7 \pm 0.8\%$, and an inclination of $i = 82.18 \pm 0.02 \, deg$. As indicated by the light curve solution, a cool starspot is placed on the primary component.

In order to study the characteristics of W UMa contact binaries, Xu-Dong Zhang and Sheng-Bang Qian (2020) have presented $a - P$, and $q - P$ relations (Equations 3 and 4)

$$a = 10.285 \times P + 0.00155 \quad (3)$$

$$log_{10}(1 + q) = 3log_{10}(10.285 \times P + 0.00155) - 2log_{10}P - log_{10}(\sqrt{5198 \times P + 2.097} - 1.481) \quad (4)$$

where $q$ is the mass ratio, $a$ is the separation between two components, and $P$ is the orbital period in year unit. The boundaries of these relations were given as,

$$a_u = 11.587 \times P + 0.00132 \quad (5)$$

$$a_l = 9.972 \times P + 0.00132 \quad (6)$$

$$log_{10}(1 + q)_u = 3log_{10}(11.587 \times P + 0.00132) - 2log_{10}P - log_{10}(\sqrt{4021 \times P + 1.868} - 1.300) \quad (7)$$

$$log_{10}(1 + q)_l = 3log_{10}(9.972 \times P + 0.00132) - 2log_{10}P - log_{10}(\sqrt{2701 \times P - 0.967} + 0.104) \quad (8)$$

We have derived $a$ and $q$ by using the above equations for BO Ari. The results of calculations the Xu-Dong Zhang and Sheng-Bang Qian (2020) relations show that the value of $a(R_\odot) = 2.256$ and its lower and upper limit are 2.149-2.451; Also, for the $q$ parameter, the value 0.3983 was obtained and the lower and upper limit of 0.0309-0.9389 was estimated; For both of these parameters, our results from this study ($a(R_\odot) = 2.152$, $q = 0.2074$) are in consistent with them. We have shown our system in Figure 6 according to these relations.

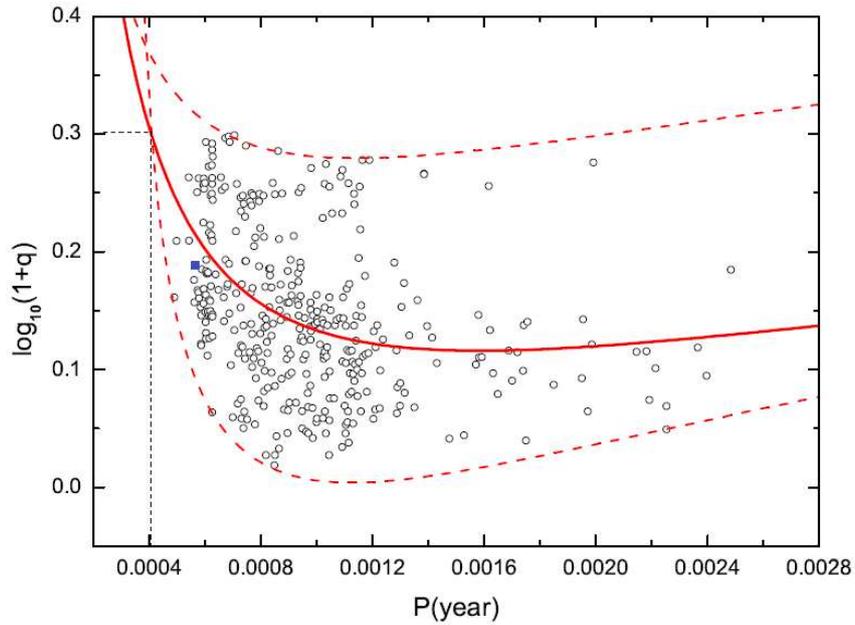

Figure 6. BO Ari's position based on the $log_{10}(1+q) - P$ relation (blue dot). The suggested value for $q$ (solid line) and the lower and upper limits for it (dotted line) are shown in the diagram.

We estimated the absolute parameters related to both components of BO Ari. Mass, radius, bolometric magnitude, and luminosity of the system were obtained. According to the estimated absolute parameters, we measured the distance as $142 \pm 9$ pc. The Gaia EDR3 parallax gives a distance value of $141.579 \pm 0.432$ pc. Therefore, our estimated distance for this binary system seems to be consistent with the Gaia EDR3 distance considering our estimated uncertainty.

The components' positions of BO Ari are plotted in the Hertzsprung-Russell (H-R) diagram and are shown in Figure 7, in which it seems the primary component is in the main-sequence, and the secondary component is placed near the ZAMS.

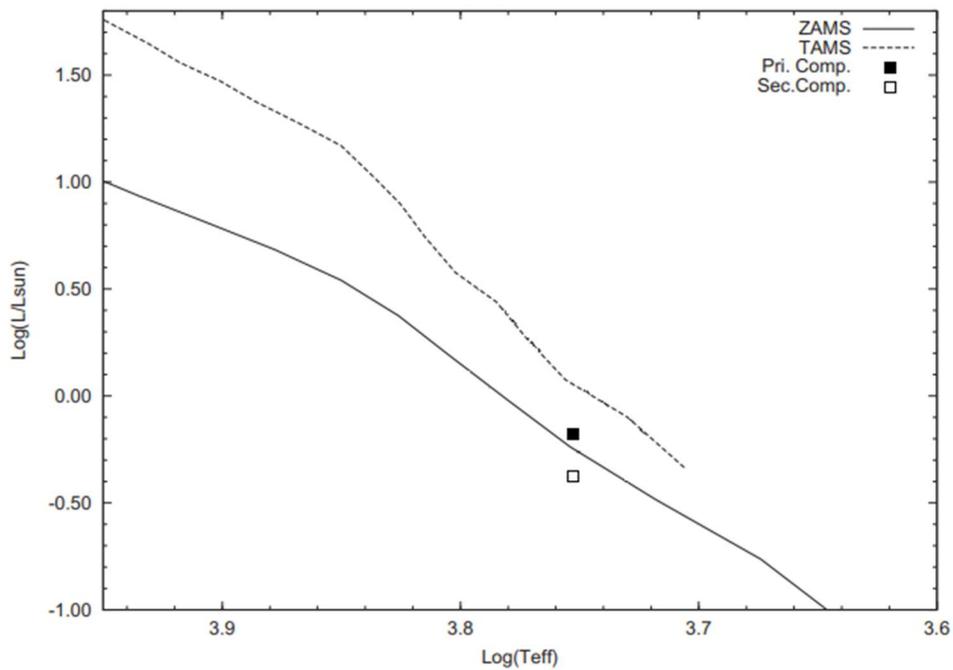

Figure 7. Position of both components of BO Ari on the H-R diagram, in which the theoretical ZAMS and TAMS curves are indicated.

According to the low mass ratio, a large of fillout factor, high inclination, and the very small temperature difference between components, we can conclude that BO Ari is an overcontact binary and A-type W UMa binary system. Given that the amount of evidence for a low mass ratio has increased from previous studies, we expect it will become deeper overcontact based on system evolution.

**ACKNOWLEDGEMENTS**
This manuscript was prepared by a joint cooperation between the International Occultation Timing Association Middle East section (IOTA/ME) and Çukurova University, Adana, Turkey. We thank TÜBİTAK National Observatory for its support in providing the CCD to UZAYMER.